# Modified Eshelby tensor for an anisotropic matrix with interfacial damage

Sangryun Lee[1,†], Jinyeop Lee[2,†], and Seunghwa Ryu[1,*]

**Affiliations**

[1]Department of Mechanical Engineering and [2]Department of Mathematical Sciences, Korea Advanced Institute of Science and Technology (KAIST), 291 Daehak-ro, Yuseong-gu, Daejeon 34141, Republic of Korea

[†]These authors contributed equally to this work.

[*]Corresponding author e-mail: ryush@kaist.ac.kr

**Abstract**

We derive a simple tensor algebraic expression of the modified Eshelby tensor for a spherical inclusion embedded in an arbitrarily anisotropic matrix in terms of three tensor quantities (the $4^{th}$ order identity tensor, the elastic stiffness tensor, and the Eshelby tensor) and two scalar quantities (the inclusion radius and interfacial spring constant), when the interfacial damage is modelled as a linear-spring layer of vanishing thickness. We validate the expression for a triclinic crystal involving 21 independent elastic constants against finite element analysis (FEA).

## 1. Introduction

The mechanical properties of multicomponent alloys and composite structures can be deduced by considering the strain fields in the inclusions and inhomogeneities. The inclusion refers to an embedded material with the identical elastic stiffness tensor $L_{pqrs}$ as the matrix, while the inhomogeneity refers to an embedded material with a different stiffness. Eshelby showed that the strain field inside the ellipsoidal inclusion embedded in an infinite matrix is uniform when the inclusion is subject to a uniform eigenstrain[1]. Eigenstrain refers to the stress-free deformation strain[2] (of the free standing inclusion) associated with thermal expansion[3], initial strain[4], or phase transformation[5-7]. The Eshelby tensor is defined as the 4$^{th}$ order tensor $S_{ijrs}$ which relates the constrained strain within the inclusion $\varepsilon_{ij}^c$ to the eigenstrain $\varepsilon_{rs}^*$, as $\varepsilon_{ij}^c = S_{ijrs}\varepsilon_{rs}^*$ [2, 8].

One can also obtain the strain field within an ellipsoidal inhomogeneity subject to an external load by transforming the problem into the corresponding equivalent inclusion problem[9, 10]. Hence, the Eshelby tensor can be used to obtain the strain field inside various embedded materials (or defects) such as voids, dislocations, cracks, reinforcements, and precipitations within an infinite matrix[11]. The mean field micromechanics models such as the Mori-Tanaka method and self-consistent method utilize the Eshelby tensor to predict the effective properties of a composite by relating the average macroscopic external strain with the average internal strain field inside inhomogeneities[12-18]. Due to the importance of the Eshelby tensor, extensive studies have been devoted to derive its simplified form for various elastic symmetries of the matrix and geometries of the inclusion[19-28].

Further developments on the single or multiple inclusion (or inhomogeneity) problem have been made by considering the interfacial damage present in realistic specimen. For example, the interface between the matrix and the reinforcement of composites synthesized through an actual manufacturing process is imperfect due to interfacial roughness, chemical corrosion, or lattice mismatch at the microscopic level[29, 30]. As a first order approximation of the interfacial damage, J. Qu introduced a linear-spring layer of vanishing thickness[15, 31, 32] at the interface to represent the displacement jump between the inclusion and the matrix. In earlier works, the simplified expressions of the modified Eshelby tensor was obtained which relates the constrained strain with the eigenstrain in the presence of the nonzero interface spring compliance[13, 14, 31, 33], and was applied to predict the effective stiffness of composites with interfacial damage[15, 16]. They considered the volume averaged constrained strain because the strain field inside the inclusion is known to be non-uniform except for in a special case[11, 31]. Later, Othmani et al. [34] pointed out a mathematical error in the earlier studies[13, 31] where the order of the surface integral and volume integral was exchanged. Such a permutation violates the Fubini-Tonelli theorem[35] because the second derivative of Green's function has a singularity within the integral domain[34, 36]. The singularity of the modified Eshelby tensor reported in earlier works[13] was caused by the violation of the Fubini-Tonelli theorem[35].

Reflecting the elastic anisotropy in the inclusion problem is also crucial for a realistic description of many technologically important materials. For instance, most polymeric products made by the extrusion process have elastic anisotropy (e.g., transversely isotropic or orthotropic) due to the partial alignment of polymer chains along the extrusion direction[37]. Multicomponent alloys involving low symmetry crystal structures are inherently anisotropic and have more than two independent elastic constants[38, 39]. However, most existing studies that consider the anisotropic matrix do not account for the interfacial damage[19, 40, 41].

Here, we derive a simple tensor algebraic expression of the modified Eshelby tensor for a spherical inclusion embedded in an arbitrarily anisotropic matrix, when the interfacial damage is modelled as a linear-spring layer of vanishing thickness. The modified Eshelby tensor expression is composed of three tensor quantities (the 4th order identity tensor, the elastic stiffness tensor, and the Eshelby tensor) and two scalar quantities (the inclusion radius and interfacial spring constant). We theoretically predict the modified Eshelby tensor of a triclinic crystal material, $NaAlSi_3O_8$, with 21 independent elastic constants and validate our results against the numerical results obtained from FEA. We show that, however, such expression cannot be obtained if the inclusion is non-spherical or if the tangential and normal compliance are not identical because the strain field inside the inclusion becomes non-uniform.

## 2. Theory

### 2.1 Eshelby tensor for the perfect interface

We introduce the governing equation for Green's function $G_{ij}(x - y)$ in elastostatics, which implies a displacement in the $i^{th}$ direction at point $x$ by the unit body force in the $j^{th}$ direction at point $y$:

$$L_{ijkl} G_{kp,lj}(x - y) + \delta_{ip}\delta(x - y) = 0 \tag{1}$$

, where $L_{ijkl}$ is the 4th order elastic stiffness tensor, and the repeated indices represent the summation over all values from 1 to 3. Green's function for the isotropic material is available in the closed form as follows:

$$G_{ij}(x - y) = \frac{1}{16\pi\mu(1-\nu)|x-y|}\left[(3-4\nu)\delta_{ij} + \frac{(x_i - y_i)(x_j - y_j)}{|x-y|^2}\right] \tag{2}$$

where $\mu$ and $\nu$ are the shear modulus and the Poisson's ratio of the material, respectively, and $|x - y|$ is the norm of vector $x - y$. For anisotropic materials, the closed form solutions were derived for transversely isotropic and cubic materials[2, 42].

The schematic for the single inclusion problem is depicted in Fig. 1, which is solved with a four step procedure. We assume that the inclusion can deform by the eigenstrain $\boldsymbol{\varepsilon}^*$ when there is no external displacement or load (step 1). In order to maintain the original shape, the load $T$ is applied to the inclusion (step 2). Then, the inclusion is plugged into a hole having the original shape and size within an infinite matrix (step 3). After removing the applied load ($T$), the inclusion exerts a traction of $F = -T$ on the matrix. Due to the constraining effect of the matrix, the inclusion deforms by the constrained strain $\boldsymbol{\varepsilon}^c$ which is different from the eigenstrain $\boldsymbol{\varepsilon}^*$, and the constrained strain field can be expressed by employing the Green's function as follows:

$$u_i^c(\boldsymbol{x}) = \int_{\partial\Omega} G_{ip}(\boldsymbol{x} - \boldsymbol{y}) F_p(\boldsymbol{y}) d\boldsymbol{y} = \int_{\partial\Omega} G_{ip}(\boldsymbol{x} - \boldsymbol{y}) \sigma_{pq}^* n_q(\boldsymbol{y}) d\boldsymbol{y} \qquad (3)$$

, where $\sigma_{pq}^* (= L_{pqrs}\varepsilon_{rs}^*)$ is called eigenstress, and $\partial\Omega$ represents the surface of the inclusion.

By applying divergence theorem, Eq. (3) can be written as

$$\varepsilon_{ij}^c(\boldsymbol{x}) = \frac{1}{2}\left(\frac{\partial u_i^c}{\partial x_j} + \frac{\partial u_j^c}{\partial x_i}\right) = \frac{1}{2}\int_\Omega L_{pqrs}\left\{\frac{\partial^2 G_{ip}(\boldsymbol{x} - \boldsymbol{y})}{\partial x_j \partial y_q} + \frac{\partial^2 G_{jp}(\boldsymbol{x} - \boldsymbol{y})}{\partial x_i \partial y_q}\right\} d\boldsymbol{y}\, \varepsilon_{rs}^*. \qquad (4)$$

When the shape of the inclusion is an ellipsoid, the integral does not depends on $\boldsymbol{x}$ within the inclusion as shown by Eshelby[1]. Hence, the equation becomes a double contraction of the constant Eshelby tensor and eigenstrain for $\boldsymbol{x} \in \Omega$. Then, we obtain

$$\varepsilon_{ij}^c = S_{ijrs}\varepsilon_{rs}^*, \text{ where } S_{ijrs} = \frac{1}{2}\int_\Omega L_{pqrs}\left\{\frac{\partial^2 G_{ip}(\boldsymbol{x} - \boldsymbol{y})}{\partial x_j \partial y_q} + \frac{\partial^2 G_{jp}(\boldsymbol{x} - \boldsymbol{y})}{\partial x_i \partial y_q}\right\} d\boldsymbol{y}. \qquad (5)$$

Because the eigenstrain and constrained strain are symmetric, the Eshelby tensor has minor symmetry ($S_{ijkl} = S_{jikl} = S_{ijlk}$). The closed form of the Eshelby tensor for anisotropic materials is available for cubic and transversely isotropic materials with a specific inclusion shape[2, 43, 44]. In order to calculate the Eshelby tensor for an anisotropic material having a symmetry lower than the transversely isotropic one, we numerically calculate the Eshelby

tensor using the formula proposed by T. Mura[2]:

$$S_{ijrs} = \frac{1}{8\pi} L_{pqrs} \int_{-1}^{1} \int_{0}^{2\pi} \{g_{ipjq}(\bar{\xi}) + g_{jpiq}(\bar{\xi})\} d\theta \, d\bar{\zeta}_3 \tag{6}$$

, where $g_{ijkl}(\bar{\xi}) = \bar{\xi}_k \bar{\xi}_l Z_{ij}(\bar{\xi})$. Here, $Z(\xi) = [(L \cdot \xi) \cdot \xi]^{-1}$, and $\xi$ are Green's function and a vector in the Fourier space, respectively. Because $Z_{ij}(\xi)$ is a homogeneous function of degree $-2$, $\xi_k \xi_l Z_{ij}(\xi)$ is identical to $\bar{\xi}_k \bar{\xi}_l Z_{ij}(\bar{\xi})$, where $\bar{\xi}$ is a normalized vector. Eq. (6) can be evaluated by using the integral variable $\bar{\zeta}_3$ and $\theta$; thus, $\bar{\xi}_1 = \frac{1}{a_1}\sqrt{1-\bar{\zeta}_3^{\,2}}\cos\theta$, $\bar{\xi}_2 = \frac{1}{a_2}\sqrt{1-\bar{\zeta}_3^{\,2}}\sin\theta$, and $\bar{\xi}_1 = \frac{1}{a_3}\bar{\zeta}_3$ with the semi-axes of the ellipsoidal inclusion $(a_i)$.

## 2.2 Interface spring model

Interfacial damage has been accounted for with a few representative methods including the interfacial spring model[11, 15, 31, 45], interphase model[46-48], and interface stress model[49-51]. The interphase model describes the interface as another phase with a finite layer thickness, and the interface stress model describes the interfacial damage using the interface stress. The interphase model adapts a third phase region along the interface with the appropriate interphase thickness and material properties, which introduce some ambiguities in describing the interfacial damage. The interface stress model assumes a coherent interface that does not account for the displacement jump. In comparison, the linear spring model does not suffer from these problems and has been applied to a variety of composite problems[11, 12, 14-16, 18, 31, 33, 34, 45] despite of its drawback, the unphysical interpenetration between the matrix and the reinforcement.

In this work, we adopt the interface spring model to consider the interfacial damage, depicted in Fig. 2. A displacement jump occurs at the interface due to the spring layer having a vanishing thickness between the infinite matrix and the single inclusion. The spring compliance is represented by $\alpha$ and $\beta$ for the tangential and normal directions, respectively, and expressed by Eq. (7) in the form of a second order tensor as follows:

$$\eta_{ij} = \alpha \delta_{ij} + (\beta - \alpha) n_i n_j \qquad (7)$$

, where $\boldsymbol{n}$ is the unit outward normal vector at the inclusion surface. The constitutive equations and traction equilibrium equation at the interface are expressed as follows:

$$\Delta t_i = \Delta \sigma_{ij} n_j = [\sigma_{ij}(\partial\Omega^+) - \sigma_{ij}(\partial\Omega^-)] n_j = 0$$

$$\Delta u_i = [u_i(\partial\Omega^+) - u_i(\partial\Omega^-)] = \eta_{ij} \sigma_{jk} n_k \qquad (8)$$

, where $(\partial\Omega^+)$ and $(\partial\Omega^-)$ denote the interface on the matrix and inclusion side, respectively. The finite spring compliances allow slip(tangential) and debonding(normal) between matrix and inclusion, but at the same time beget unphysical overlapping under compressive traction. An in-depth discussion on this issue is beyond the scope of this paper, and we refer to a previous work concerning the problem[52]. We also note that the spring compliances are assumed to be constant over the entire range of separation distance. Hence, the present model is inadequate to describe the fracture behavior of the composites concerning interfacial failure, while it is applicable to characterize the elastic response. Formulating the Eshelby inclusion problem by adopting the interfacial condition in Eq. (8), the constrained strain is written as

$$\varepsilon_{ij}^c = S_{ijrs} \varepsilon_{rs}^* + \frac{1}{2} L_{klmn} L_{pqrs} \int_{\partial\Omega} \eta_{kp} \left\{ \frac{\partial^2 G_{im}(\boldsymbol{x}-\boldsymbol{y})}{\partial x_j \partial y_n} \right.$$

$$\left. + \frac{\partial^2 G_{jm}(\boldsymbol{x}-\boldsymbol{y})}{\partial x_i \partial y_n} \right\} n_q(\boldsymbol{y}) n_l(\boldsymbol{y}) (\varepsilon_{rs}^c(\boldsymbol{y}) - \varepsilon_{rs}^*) d\boldsymbol{y}. \qquad (9)$$

As shown in Eq. (9), it reproduces the perfect interface case with zero spring compliance, i.e. $\eta_{ij} = 0$.

Because Eq. (9) is an implicit integral equation involving the constrained strain $\varepsilon_{ij}^c$, it is difficult to obtain the relation between $\varepsilon_{ij}^c$ and $\varepsilon_{rs}^*$, i.e., the modified Eshelby tensor. Zhong et al. showed that, when $\alpha = \beta \equiv \gamma$, the strain field inside the spherical inclusion is uniform as in the case of the perfect interface[11]. Hence, for this special case, the integral equation can be decomposed as follows:

$$\varepsilon_{ij}^c = S_{ijrs}\varepsilon_{rs}^* - \Gamma_{ijrs}(\varepsilon_{rs}^c - \varepsilon_{rs}^*) \tag{10}$$

, where the 4$^{th}$ order tensor $\Gamma_{ijkl}$ in Eq. (10) is defined by

$$\Gamma_{ijrs} \equiv -\frac{1}{2}\gamma L_{plmn}L_{pqrs}\int_{\partial\Omega}\left\{\frac{\partial^2 G_{im}(x-y)}{\partial x_j \partial y_n} + \frac{\partial^2 G_{jm}(x-y)}{\partial x_i \partial y_n}\right\}n_q(y)n_l(y)dy. \tag{11}$$

The constrained strain field then can be expressed in a tensor algebraic equation as follows:

$$\boldsymbol{\varepsilon}^c = (\boldsymbol{I} + \boldsymbol{\Gamma})^{-1}:(\boldsymbol{S} + \boldsymbol{\Gamma}):\boldsymbol{\varepsilon}^* = \boldsymbol{S}^M:\boldsymbol{\varepsilon}^* \tag{12}$$

, where the colon ":" denotes a double contraction; $\boldsymbol{I}$ is the 4$^{th}$ order symmetric identity tensor such that $I_{ijkl} = \frac{1}{2}(\delta_{ik}\delta_{jl} + \delta_{il}\delta_{jk})$, and $\boldsymbol{S}^M$ is the modified Eshelby tensor. For isotropic materials, by plugging the Green's function expression in Eq. (2) into Eq. (11), $\Gamma_{ijkl}$ can be expressed as follows:

$$\Gamma_{ijkl} = \frac{\mu\gamma}{15R(1-\nu)}[2(1+5\nu)\delta_{ij}\delta_{kl} + (7-5\nu)(\delta_{ik}\delta_{jl} + \delta_{il}\delta_{jk})], \tag{13}$$

, which already was shown by Y. Othmani et al.[34], where $R$ is the radius of the inclusion. For the ellipsoidal inclusion case, Eq. (10) is not valid because the constrained strain field within the inclusion is non-uniform. We carried out numerical calculations to confirm that the constrained strain field within the inclusion is non-uniform (See appendix 1). For the anisotropic spherical inclusion, it is difficult to calculate $\Gamma_{ijkl}$ in an explicit form because the closed form of Green's function is not available in general.

Here, without using the Green's function expression, we prove that, for materials having spherical inclusion with an arbitrary elastic anisotropy, the $\boldsymbol{\Gamma}$ tensor can be written in terms of the Eshelby tensor for the perfect interface and the elastic stiffness tensor. By using the relation $\frac{\partial^2 G_{im}(x-y)}{\partial x_j \partial y_n} = -\frac{\partial^2 G_{im}(x-y)}{\partial y_j \partial y_n}$ and applying the divergence theorem to Eq. (11), we obtain

$$\Gamma_{ijrs} = \frac{1}{2}\gamma L_{plmn}L_{pqrs}\int_\Omega \left[\left\{\frac{\partial^3 G_{im}(x-y)}{\partial y_l \partial y_j \partial y_n} + \frac{\partial^3 G_{jm}(x-y)}{\partial y_l \partial y_i \partial y_n}\right\}n_q(y)\right.$$
$$\left. + \left\{\frac{\partial^2 G_{im}(x-y)}{\partial y_j \partial y_n} + \frac{\partial^2 G_{jm}(x-y)}{\partial y_i \partial y_n}\right\}\frac{\partial n_q}{\partial y_l}\right]dy. \tag{14}$$

Because $n_q(y) = y_q/R$ for the spherical inclusion, Eq. (14) leads to

$$\Gamma_{ijrs} = \frac{\gamma}{R}L_{pqrs}\left[\frac{1}{2}L_{plmn}\int_\Omega \left\{\frac{\partial^3 G_{im}(x-y)}{\partial y_l \partial y_j \partial y_n} + \frac{\partial^3 G_{jm}(x-y)}{\partial y_l \partial y_i \partial y_n}\right\}y_q dy\right.$$
$$\left. + L_{mnpq}\int_\Omega \frac{1}{2}\left\{\frac{\partial^2 G_{im}(x-y)}{\partial y_j \partial y_n} + \frac{\partial^2 G_{jm}(x-y)}{\partial y_i \partial y_n}\right\}dy\right]. \tag{15}$$

In Eq. (15), the 2nd integral on the right hand side is the definition of the Eshelby tensor; thus, it can be simplified as follows:

$$\Gamma_{ijrs} = \frac{\gamma}{R}\left[\frac{1}{2}L_{plmn}\int_\Omega \left\{\frac{\partial^3 G_{im}(x-y)}{\partial y_l \partial y_j \partial y_n} + \frac{\partial^3 G_{jm}(x-y)}{\partial y_l \partial y_i \partial y_n}\right\}y_q dy - S_{ijpq}\right]L_{pqrs}. \tag{16}$$

To further simplify the integration in Eq. (16), we consider the following equation which is obtained by multiplying $y_q$ after differentiating Eq. (1) with respect to $y_j$,

$$L_{plmn}\frac{\partial^3 G_{mi}(x-y)}{\partial y_j \partial y_l \partial y_n}y_q = -\frac{\partial}{\partial y_j}\left(\delta_{ip}\delta(x-y)\right)y_q. \tag{17}$$

When the divergence theorem is used after integrating Eq. (17) for the inclusion volume, it leads to Eq. (18), and eventually, the integral in Eq. (16) is reduced as the 4th order identity tensor because

$$\int_\Omega L_{plmn} \frac{\partial^3 G_{mi}(x-y)}{\partial y_j \partial y_l \partial y_n} y_q d\boldsymbol{y} = \int_\Omega \left(\delta_{ip}\delta(x-y)\right)\frac{\partial y_q}{\partial y_j} d\boldsymbol{y} = \delta_{ip}\delta_{jq} \tag{18}$$

and thus

$$\frac{1}{2}L_{plmn}\int_\Omega \left\{\frac{\partial^3 G_{im}(x-y)}{\partial y_l \partial y_j \partial y_n} + \frac{\partial^3 G_{jm}(x-y)}{\partial y_l \partial y_i \partial y_n}\right\} y_q d\boldsymbol{y} = \frac{1}{2}(\delta_{ip}\delta_{jq} + \delta_{jp}\delta_{iq}) = I_{ijpq}. \tag{19}$$

Using Eq. (19), the $\boldsymbol{\Gamma}$ for anisotropic matrix can be shown as

$$\Gamma_{ijrs} = \frac{\gamma}{R}(I_{ijpq} - S_{ijpq})L_{pqrs}. \tag{20}$$

Therefore, it is proven that the modified Eshelby tensor $\boldsymbol{S}^M$ can be written as follows:

$$\boldsymbol{S}^M = \left[\boldsymbol{I} + \frac{\gamma}{R}(\boldsymbol{I}-\boldsymbol{S}):\boldsymbol{L}\right]^{-1} : \left[\boldsymbol{S} + \frac{\gamma}{R}(\boldsymbol{I}-\boldsymbol{S}):\boldsymbol{L}\right]. \tag{21}$$

## 3. Numerical Validation

### 3.1 Finite element analysis

To verify the expression for the modified Eshelby tensor, we first computed the Eshelby tensor for the perfect interface by FEA. Because the Eshelby tensor is defined for the infinite matrix, we conducted a series of FEA to check the convergence of the numerically computed Eshelby tensor by changing the size $L$ of the matrix under a fixed inclusion size $D(=2mm)$ (See Figure. 3). Due to the jump discontinuity of some of the stress and strain components across the interface, we use sufficiently fine mesh (maximum size: 0.2mm) at the interface which yields the strain distribution $\varepsilon_{11}/\varepsilon_{11}^*$ almost identical to that predicted from a much finer mesh model (See Appendix 2). The computation was performed using the COMSOL software[53], and about 400,000 quadratic tetrahedron elements were used for $L/D = 2.5, 5.0, 7.5,$ and $10.0$. The unit eigenstrain was assigned to the inclusion, and a fixed boundary condition was applied to the outer surface of the matrix (see Figure. 3(a)). In order to implement the perfect bonding assumption at the interface, the 'tie' condition was used, and the Eshelby tensor was obtained from the constrained strain. In this study, we considered the NaAlSi$_3$O$_8$[54] crystal which has 21 independent elastic constants shown in Table.1. We plotted 6 diagonal components in the Eshelby tensor ($S_{IJIJ}$, here $I$ and $J$ are not dummy index) for various $L/D$ (see Figure. 3(b)). If the edge length of the matrix is 10 times longer than the diameter of the inclusion($L > 10D$), that results from the FEA agree well with the theoretical values predicted by Eq. (6). Because the FEA results are obtained under a fixed boundary condition at the outer surface, the theoretical value can be regarded as the upper bound for the FEA results. All 36 independent components are summarized in Table.2.

**3.2 Modified Eshelby tensor**

In order to calculate the modified Eshelby tensor in Eq. (21), the inverse and double contractions of the 4$^{th}$ order tensor should be conducted. Previous studies have performed inverse or double contractions of the 4$^{th}$ order tensors which have six or less independent coefficients based on Walpole's notation[12, 18, 55]. However, Walpole's notation is not applicable for an anisotropic material with 21 elastic constants and 36 independent Eshelby tensor components. Hence, in the present study, we adopted the Mandel notation[56] to facilitate the inverse and double inner product operations of the 4$^{th}$ order tensors. In the Mandel notation, the stress vector $\vec{\sigma}$, strain vector $\vec{\varepsilon}$, stiffness matrix $\langle L \rangle$, and Eshelby matrix $\langle S \rangle$ are defined as follows:

$$\vec{\sigma} = \begin{bmatrix} \sigma_{11} \\ \sigma_{22} \\ \sigma_{33} \\ \sqrt{2}\sigma_{23} \\ \sqrt{2}\sigma_{31} \\ \sqrt{2}\sigma_{12} \end{bmatrix}, \qquad \vec{\varepsilon} = \begin{bmatrix} \varepsilon_{11} \\ \varepsilon_{22} \\ \varepsilon_{33} \\ \sqrt{2}\varepsilon_{23} \\ \sqrt{2}\varepsilon_{31} \\ \sqrt{2}\varepsilon_{12} \end{bmatrix},$$

$$\langle L \rangle = \begin{bmatrix} L_{1111} & L_{1122} & L_{1133} & \sqrt{2}L_{1123} & \sqrt{2}L_{1131} & \sqrt{2}L_{1112} \\ L_{1122} & L_{2222} & L_{2233} & \sqrt{2}L_{2223} & \sqrt{2}L_{2231} & \sqrt{2}L_{2212} \\ L_{1133} & L_{2233} & L_{3333} & \sqrt{2}L_{3323} & \sqrt{2}L_{3331} & \sqrt{2}L_{3312} \\ \sqrt{2}L_{1123} & \sqrt{2}L_{2223} & \sqrt{2}L_{3323} & 2L_{2323} & 2L_{2331} & 2L_{2312} \\ \sqrt{2}L_{1131} & \sqrt{2}L_{2231} & \sqrt{2}L_{3331} & 2L_{2331} & 2L_{3131} & 2L_{3112} \\ \sqrt{2}L_{1112} & \sqrt{2}L_{2212} & \sqrt{2}L_{3312} & 2L_{2312} & 2L_{3112} & 2L_{1212} \end{bmatrix} \qquad (22)$$

$$\langle S \rangle = \begin{bmatrix} S_{1111} & S_{1122} & S_{1133} & \sqrt{2}S_{1123} & \sqrt{2}S_{1131} & \sqrt{2}S_{1112} \\ S_{2211} & S_{2222} & S_{2233} & \sqrt{2}S_{2223} & \sqrt{2}S_{2231} & \sqrt{2}S_{2212} \\ S_{3311} & S_{3322} & S_{3333} & \sqrt{2}S_{3323} & \sqrt{2}S_{3331} & \sqrt{2}S_{3312} \\ \sqrt{2}S_{2311} & \sqrt{2}S_{2322} & \sqrt{2}S_{2333} & 2S_{2323} & 2S_{2331} & 2S_{2312} \\ \sqrt{2}S_{3111} & \sqrt{2}S_{3122} & \sqrt{2}S_{3133} & 2S_{3123} & 2S_{3131} & 2S_{3112} \\ \sqrt{2}S_{1211} & \sqrt{2}S_{1222} & \sqrt{2}S_{1233} & 2S_{1223} & 2S_{1231} & 2S_{1212} \end{bmatrix}$$

The prefactors $\sqrt{2}$ and 2 ensure that the matrix-matrix product and the inverse coincide with the double contraction and the inverse of the 4$^{\text{th}}$ order tensors, respectively. Hence, if $\mathbf{A}, \mathbf{B}$ are the 4$^{\text{th}}$ order tensor with minor symmetry ($A_{ijkl} = A_{jikl} = A_{ijlk}$), and $\langle \mathbf{A} \rangle, \langle \mathbf{B} \rangle$ are the corresponding $6 \times 6$ matrix following the Mandel notation, we can calculate the double contraction and inverse from the $6 \times 6$ matrix multiplication and inverse, respectively, as

$$\langle \mathbf{A} : \mathbf{B} \rangle = \langle \mathbf{A} \rangle \langle \mathbf{B} \rangle, \qquad \langle \mathbf{A}^{-1} \rangle = \langle \mathbf{A} \rangle^{-1}. \qquad (23)$$

Thus, we can predict the modified Eshelby tensor using the $6 \times 6$ matrix multiplication and inverse such that

$$\langle \mathbf{S}^M \rangle = \left[ \langle \mathbf{I} \rangle + \frac{\gamma}{R}(\langle \mathbf{I} \rangle - \langle \mathbf{S} \rangle)\langle \mathbf{L} \rangle \right]^{-1} \left[ \langle \mathbf{S} \rangle + \frac{\gamma}{R}(\langle \mathbf{I} \rangle - \langle \mathbf{S} \rangle)\langle \mathbf{L} \rangle \right]. \qquad (24)$$

We note that, when the frequently used Voigt notation for the matrix representation[57-60] is used, we cannot perform the tensor operations as in Eq. (23) and Eq. (24) because the transformed

$6 \times 6$ matrices of the stiffness and Eshelby tensor have different prefactors depending on whether it relates stress to the strain ($\boldsymbol{\sigma} = \boldsymbol{L} : \boldsymbol{\varepsilon}$) or strain to strain ($\boldsymbol{\varepsilon}^c = \boldsymbol{S} : \boldsymbol{\varepsilon}^*$)(See Eqn. (25)).

$$\vec{\sigma}^{\text{voigt}} = \begin{bmatrix} \sigma_{11} \\ \sigma_{22} \\ \sigma_{33} \\ \sigma_{23} \\ \sigma_{31} \\ \sigma_{12} \end{bmatrix}, \qquad \vec{\varepsilon}^{\text{voigt}} = \begin{bmatrix} \varepsilon_{11} \\ \varepsilon_{22} \\ \varepsilon_{33} \\ 2\varepsilon_{23} \\ 2\varepsilon_{31} \\ 2\varepsilon_{12} \end{bmatrix},$$

$$\langle \boldsymbol{L} \rangle^{\text{voigt}} = \begin{bmatrix} L_{1111} & L_{1122} & L_{1133} & L_{1123} & L_{1131} & L_{1112} \\ L_{1122} & L_{2222} & L_{2233} & L_{2223} & L_{2231} & L_{2212} \\ L_{1133} & L_{2233} & L_{3333} & L_{3323} & L_{3331} & L_{3312} \\ L_{1123} & L_{2223} & L_{3323} & L_{2323} & L_{2331} & L_{2312} \\ L_{1131} & L_{2231} & L_{3331} & L_{2331} & L_{3131} & L_{3112} \\ L_{1112} & L_{2212} & L_{3312} & L_{2312} & L_{3112} & L_{1212} \end{bmatrix} \qquad (25)$$

$$\langle \boldsymbol{S} \rangle^{\text{voigt}} = \begin{bmatrix} S_{1111} & S_{1122} & S_{1133} & S_{1123} & S_{1131} & S_{1112} \\ S_{2211} & S_{2222} & S_{2233} & S_{2223} & S_{2231} & S_{2212} \\ S_{3311} & S_{3322} & S_{3333} & S_{3323} & S_{3331} & S_{3312} \\ 2S_{2311} & 2S_{2322} & 2S_{2333} & 2S_{2323} & 2S_{2331} & 2S_{2312} \\ 2S_{3111} & 2S_{3122} & 2S_{3133} & 2S_{3123} & 2S_{3131} & 2S_{3112} \\ 2S_{1211} & 2S_{1222} & 2S_{1233} & 2S_{1223} & 2S_{1231} & 2S_{1212} \end{bmatrix}$$

We compared the theoretical predictions of the modified Eshelby tensor with the FEA results shown in Fig. 4. In the FEA, the interface spring condition at the interface was implemented, and the calculation was performed under the same mesh condition as in Section 3.1, and a $L/D$ ratio of 10 was used. All 36 independent components of the modified Eshelby tensor are well matched with the FEA for the entire range of the interfacial damage. In the limit of the infinite interface spring compliance, all the diagonal components converge to 1, and the off-diagonal components converge to zero because the constrained strain converges to the eigenstrain for infinite compliance (zero stiffness). We can also explain the finding based on the Eq. (21). When the interfacial damage $\gamma$ goes to zero, it reproduces the Eshelby tensor for the perfect interface $S$, and as $\gamma$ goes to infinite, the modified Eshelby tensor converges to the identity tensor. We computed an invariant of the modified Eshelby tensor, $S_{ijij}$, both analytically and computationally. The invariant $S_{ijij}$ is known as 3 for the perfect bonding[61] and is expected to be 6 in the limit of the infinite spring compliance because the modified Eshelby tensor converges to the identity tensor (See Eq. (21)). Fig. 5 shows the invariant bounds between the two limiting values of 3 and 6 for three different materials ($NaAlSi_3O_8$, ammonium tetroxalate dehydrate (ATO), and parallelepiped Si) having triclinic symmetry[54, 62, 63] (See Fig. 5). The elastic constants of the ATO and parallelepipied Si are listed in Table. 2.

## 4. Conclusion

We derive a simple tensor algebraic expression for the modified Eshelby tensor for materials with an arbitrary elastic anisotropy when the interfacial damage is described by the interfacial spring model. Once the Eshelby tensor for the perfect interfacial is obtained analytically or numerically, the modified Eshelby tensor can be obtained analytically by a few matrix operations. We validated our theoretical prediction against FEA results for the entire range of interfacial damage and show that the invariant in the modified Eshelby tensor is bounded between 3 and 6. Our finding can be applied to a wide range of composite problems involving interfacial damage and matrices with an elastic anisotropy.

**Acknowledgements**

This work is supported by the National Research Foundation of Korea (NRF) funded by the Ministry of Science and ICT (2016M3D1A1900038 and 2016R1C1B2011979).

## Appendix 1 Ellipsoidal inclusion case

In this section, we prove that the ellipsoidal inclusion has a non-uniform strain field when interfacial damage exists. We simulate the eigenstrain problem when the inclusion shape is ellipsoidal. The semi-axes of the inclusion is $2\text{mm}, 1\text{mm}, 1\text{mm}$ in the $x, y, z$ directions, respectively, and the length of the matrix domain is ten times larger than each axis length of the inclusion ($20\text{mm} \times 10\text{mm} \times 10\text{mm}$). The material properties used in the FEA and numerical calculation are $E = 200\text{GPa}$ and $\nu = 0.25$. As proved by Eshelby, the constrained strain field is uniform for the perfect bonding case, but unlike the spherical inclusion case the strain field within the inclusion is non-uniform when interfacial damage exists, even for small interfacial damage (See Figure. 6).

To explain the non-uniform strain field, we calculate $M_{111111}$ with respect to the $x$ axis of the ellipsoidal for different aspect ratios.

$$M_{ijmnql} \equiv \frac{1}{2} \int_{\partial \Omega} \left\{ \frac{\partial^2 G_{im}(\boldsymbol{x} - \boldsymbol{y})}{\partial x_j \partial y_n} + \frac{\partial^2 G_{jm}(\boldsymbol{x} - \boldsymbol{y})}{\partial x_i \partial y_n} \right\} n_q(\boldsymbol{y}) n_l(\boldsymbol{y}) d\boldsymbol{y}. \tag{A1}$$

As shown in Fig. 6(a), the $M_{111111}$ is constant for the spherical inclusion case (aspect ratio=1). However, the $M_{111111}$ depends on position $\boldsymbol{x}$ when the aspect ratio is less or larger than 1. Thus, we can explain that the Eq. (10) is not valid for the ellipsoidal case because the strain field ($\boldsymbol{\varepsilon}^c$) within the inclusion is not uniform.

## Appendix 2. Meshing in the Finite Element Analyses

We construct 409,591 elements on matrix and 8,718 elements on inclusion for all FEA calculations presented in this study, as depicted in Figure 7. The maximum size of elements in inclusion is set as 0.2 mm. To demonstrate the mesh convergence with a specific example, we consider the strain distribution $\varepsilon_{xx}/\varepsilon_{xx}^*$ along the radial direction for the NaAlSi$_3$O$_8$ matrix having a spherical inclusion with interfacial spring compliance $\gamma = 10^{-2}$ mm/GPa, when $\varepsilon_{xx}^* > 0$ and other eigenstrain components are zero. As shown in Figure 7, the strain within and outside of the inclusion predicted from our model matches well with the strain predicted from a much finer mesh model.

**Figures and captions**

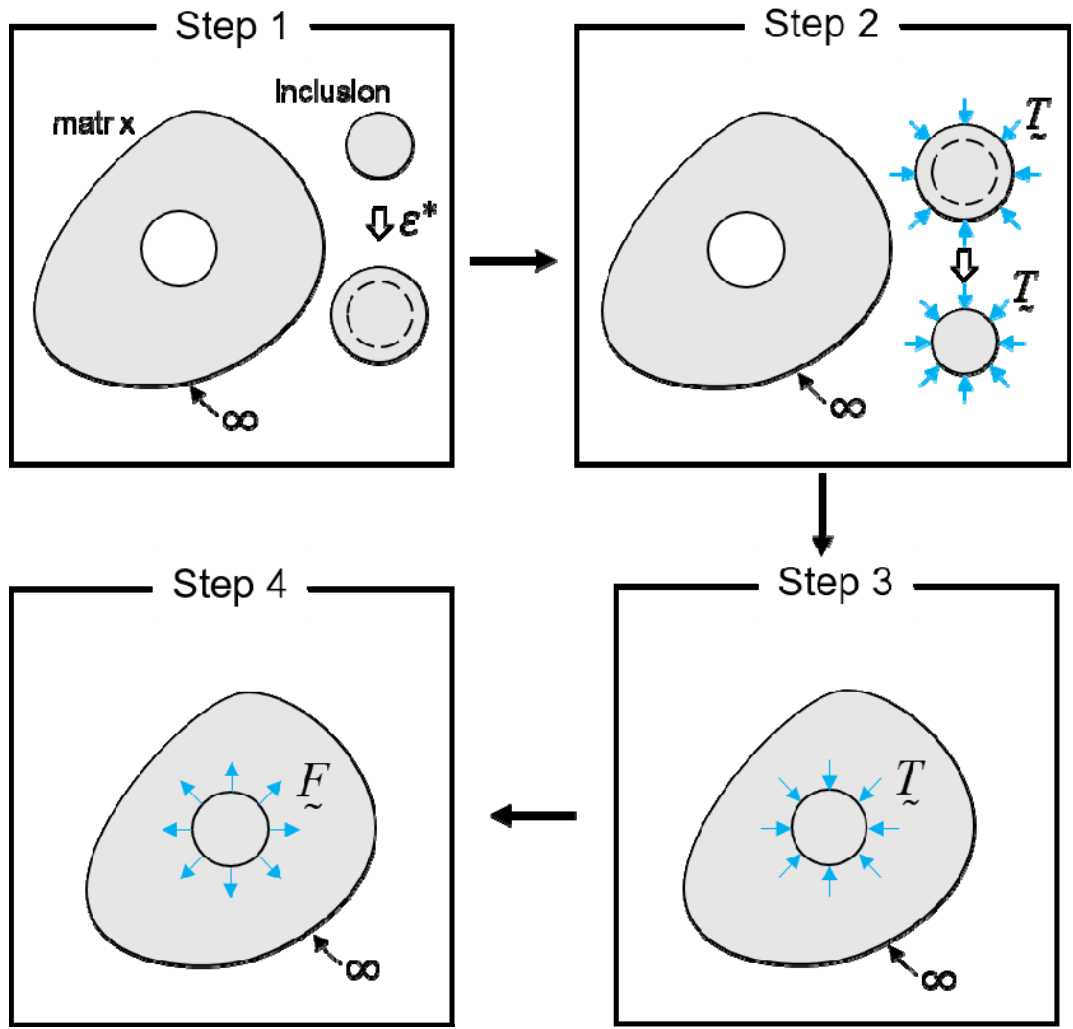

**Figure 1**. Schematic of the single inclusion problem.

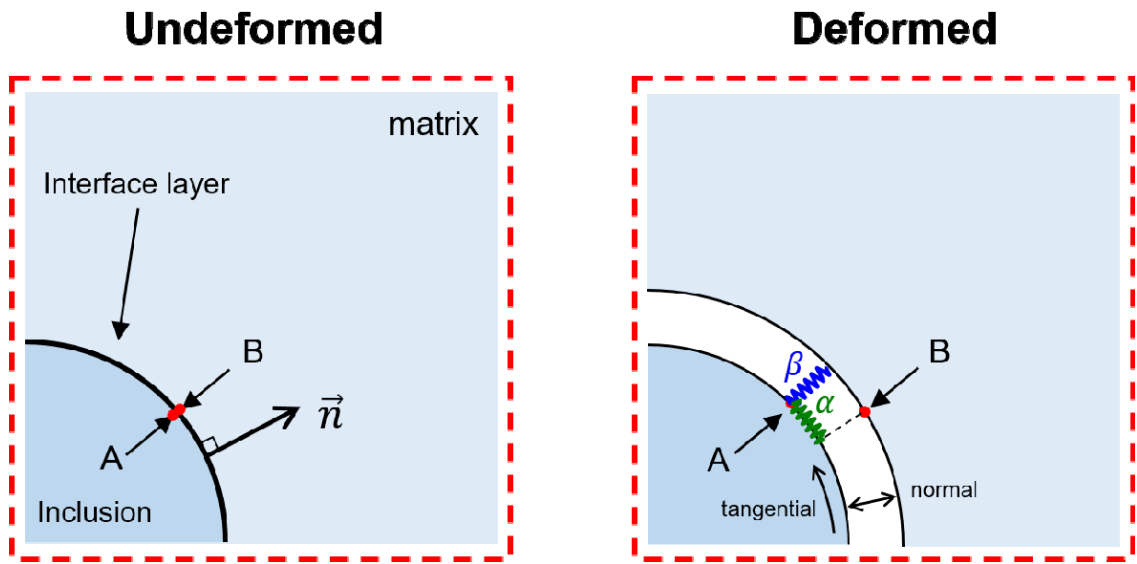

**Figure 2**. Schematic of the interface spring model for undeformed and deformed state.

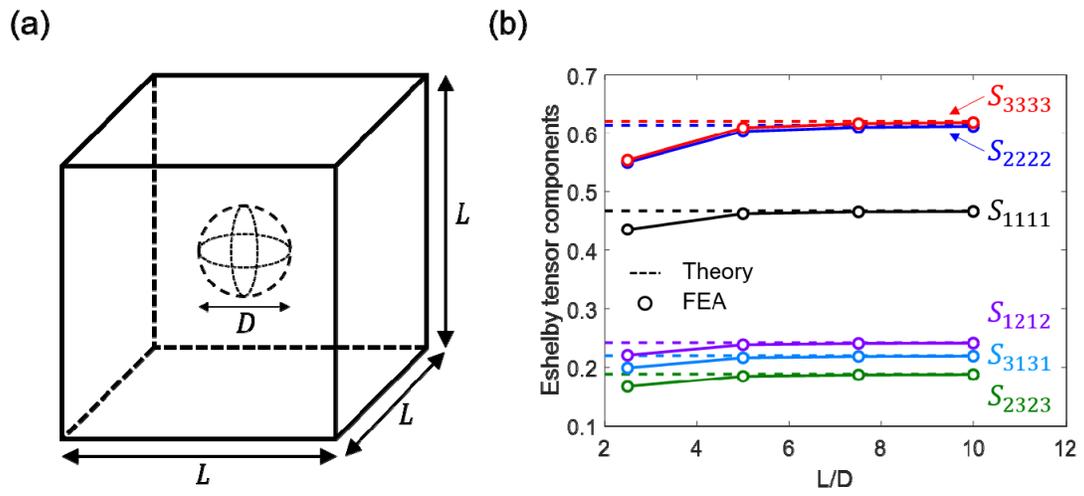

**Figure 3**. (a) Geometry used for the FEA. (b) Eshelby tensor components ($S_{IJIJ}$) for $L/D$ in the absence of interfacial damage.

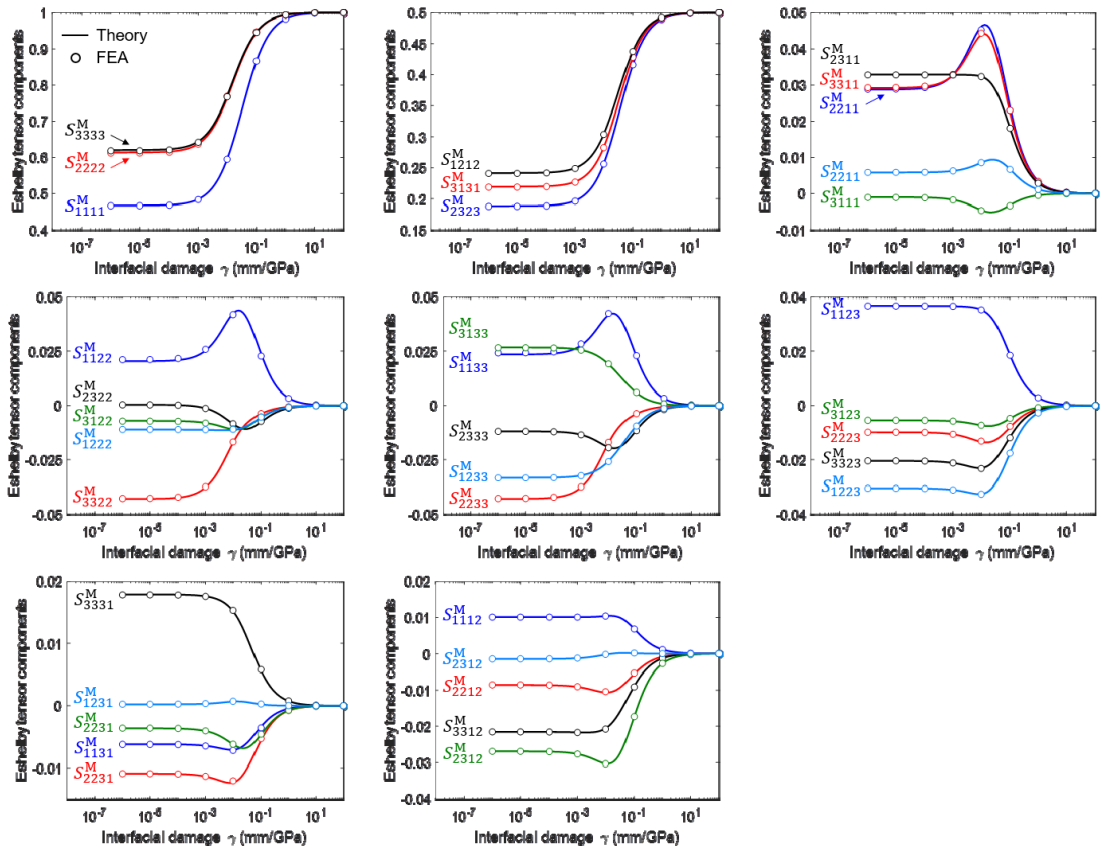

**Figure 4**. All 36 independent modified Eshelby tensor components with respect to interfacial spring compliance for the triclinic crystal NaAlSi$_3$O$_8$ with 21 independent elastic constants.

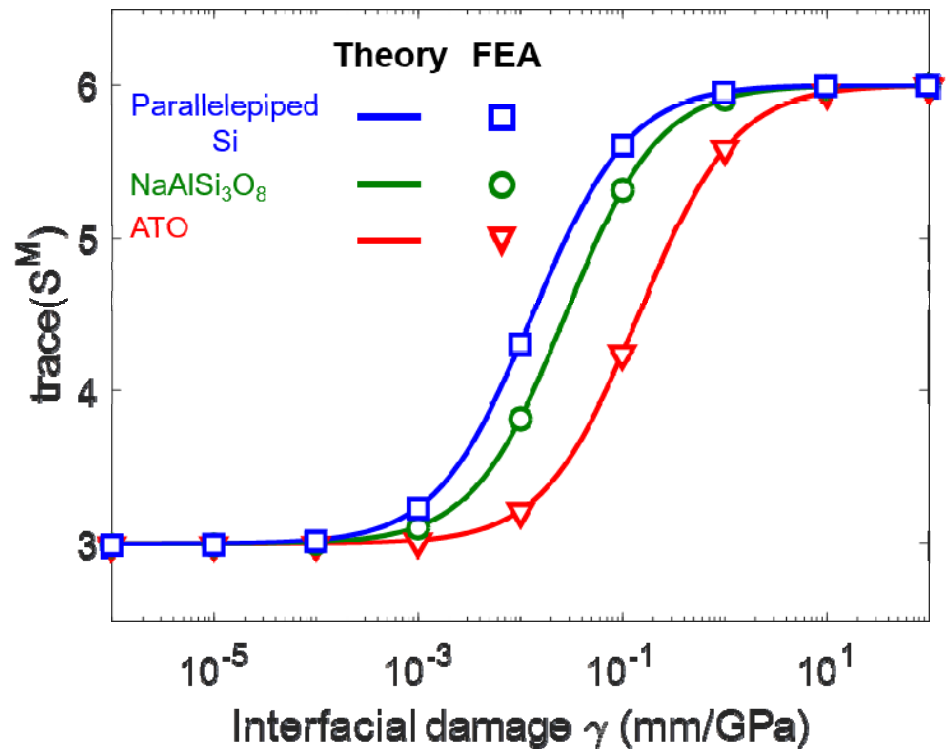

**Figure 5.** An invariant ($S_{ijij}$) of the modified Eshelby tensor for different triclinic crystals.

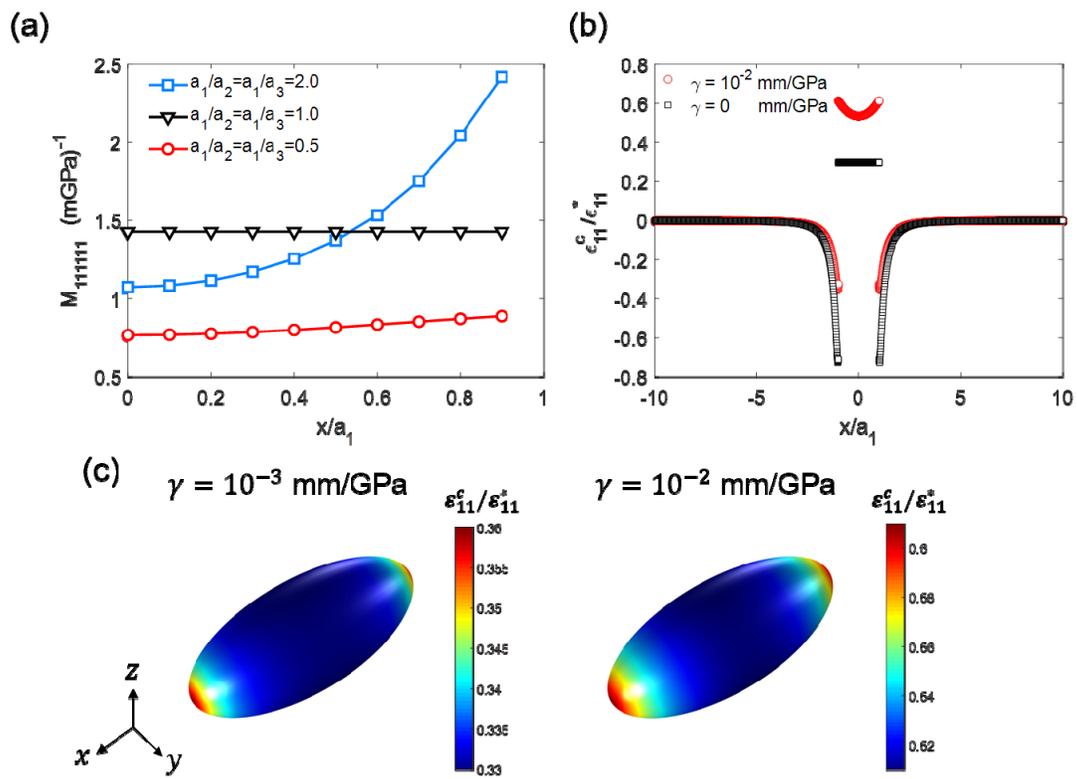

**Figure 6**. (a) Numerical prediction of $M_{111111}$ for different ellipsoidal inclusions with respect to *x*. (b) Constrained strain field within the inclusion for the perfect bonding and imperfect bonding case. (c) 3D plot of the constrained strain field for two different interfacial damages.

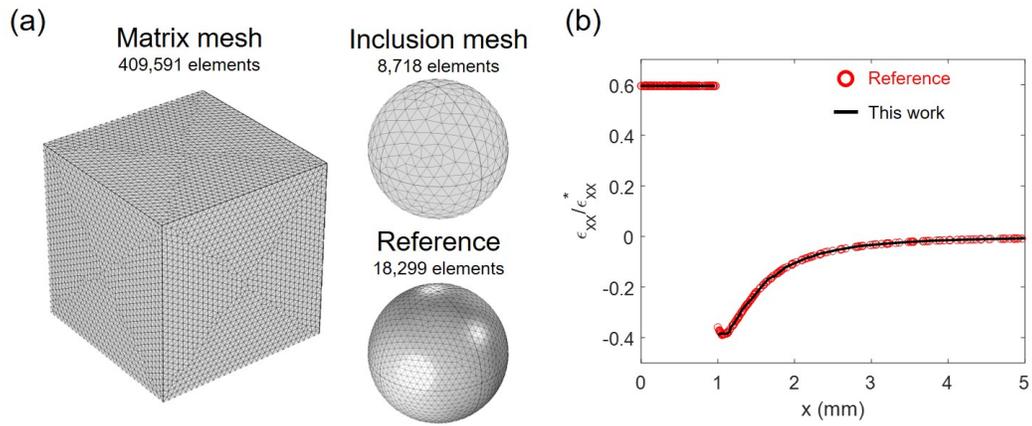

**Figure 7**. (a) Our mesh model of the matrix and inclusion, and a reference model with a much finer mesh for the mesh convergence test. Inclusion is arbitrarily magnified for visualization purpose. (b) Predicted $\varepsilon_{xx}/\varepsilon_{xx}^*$ as a function of radial distance $x$ ($x = 0$ is the center of the inclusion.) for the NaAlSi$_3$O$_8$ matrix having a spherical inclusion with interfacial spring compliance $\gamma = 10^{-2}$ mm/GPa, when $\varepsilon_{xx}^* > 0$ and other eigenstrain components are zero.

**Table.1** Elastic stiffness ($L_{ijkl}$ in GPa) of NaAlSi$_3$O$_8$, ammonium tetroxalate dehydrate (ATO) and parallelepiped Si.

| (i,j) \ (k,l) | (1,1) | (2,2) | (3,3) | (2,3) | (3,1) | (1,2) |
|---|---|---|---|---|---|---|
| **NaAlSi$_3$O$_8$** | | | | | | |
| (1,1) | 69.1 | 34 | 30.8 | 5.1 | -2.4 | -0.9 |
| (2,2) | | 183.5 | 5.5 | -3.9 | -7.7 | -5.8 |
| (3,3) | | | 179.5 | -8.7 | 7.1 | -9.8 |
| (2,3) | | | | 24.9 | -2.4 | -7.2 |
| (3,1) | | | | | 26.8 | 0.5 |
| (1,2) | | | | | | 33.5 |
| **ATO** | | | | | | |
| (1,1) | 21.87 | 11.99 | 10.39 | 1.63 | 5.99 | -1.03 |
| (2,2) | | 45.89 | 16.29 | 11.56 | 2.02 | -3.77 |
| (3,3) | | | 36.38 | 3.77 | 2.03 | -0.76 |
| (2,3) | | | | 10.43 | 0.14 | 0.15 |
| (3,1) | | | | | 5.40 | 0.12 |
| (1,2) | | | | | | 4.44 |
| **Parallelepiped Si** | | | | | | |
| (1,1) | 197.41 | 56.86 | 39.81 | -6.52 | -6.75 | 4.11 |
| (2,2) | | 183.65 | 54.21 | 9.65 | 8.04 | -8.46 |
| (3,3) | | | 199.63 | -7.75 | -1.45 | 5.48 |
| (2,3) | | | | 70.23 | 4.46 | 8.18 |
| (3,1) | | | | | 55.58 | -5.46 |
| (1,2) | | | | | | 73.07 |

**Table.2** All independent Eshelby tensor components ($S_{ijkl}$) for the perfect bonding problem.

| Theory | | | | | | |
|---|---|---|---|---|---|---|
| (k,l) \ (i,j) | (1,1) | (2,2) | (3,3) | (2,3) | (3,1) | (1,2) |
| (1,1) | 0.467491 | 0.023939 | 0.022871 | 0.073494 | -0.012146 | 0.020443 |
| (2,2) | 0.030553 | 0.606584 | -0.045073 | -0.020579 | -0.023196 | -0.017802 |
| (3,3) | 0.028397 | -0.042124 | 0.623481 | -0.039842 | 0.035215 | -0.042064 |
| (2,3) | 0.033206 | -0.000214 | -0.011539 | 0.378851 | -0.007503 | -0.053832 |
| (3,1) | -0.000987 | -0.008043 | 0.026333 | -0.010817 | 0.438151 | -0.002995 |
| (1,2) | 0.006075 | -0.011390 | -0.033238 | -0.061113 | 0.000388 | 0.485440 |
| **Finite element analysis** | | | | | | |
| (k,l) \ (i,j) | (1,1) | (2,2) | (3,3) | (2,3) | (3,1) | (1,2) |
| (1,1) | 0.466633 | 0.024617 | 0.023620 | 0.073362 | -0.012204 | 0.020423 |
| (2,2) | 0.030669 | 0.604826 | -0.044916 | -0.020558 | -0.023220 | -0.017727 |
| (3,3) | 0.028508 | -0.041929 | 0.621449 | -0.039746 | 0.035085 | -0.042056 |
| (2,3) | 0.033103 | -0.000257 | -0.011560 | 0.377259 | -0.007438 | -0.053814 |
| (3,1) | -0.001046 | -0.008166 | 0.026211 | -0.010782 | 0.436733 | -0.002922 |
| (1,2) | 0.006095 | -0.011313 | -0.033303 | -0.061150 | 0.000432 | 0.484091 |

# Reference

1. Eshelby JD. The determination of the elastic field of an ellipsoidal inclusion, and related problems. *Proceedings of the Royal Society of London Series A Mathematical and Physical Sciences* 1957; 241: 376. 10.1098/rspa.1957.0133.
2. Mura T. *Micromechanics of defects in solids*. Springer Science & Business Media, 2013.
3. Jun TS and Korsunsky AM. Evaluation of residual stresses and strains using the Eigenstrain Reconstruction Method. *Int J Solids Struct* 2010; 47: 1678-1686. DOI: 10.1016/j.ijsolstr.2010.03.002.
4. Chiu YP. On the Stress Field Due to Initial Strains in a Cuboid Surrounded by an Infinite Elastic Space. *Journal of Applied Mechanics* 1977; 44: 587-590. DOI: 10.1115/1.3424140.
5. Dvorak GJ and Benveniste Y. On Transformation Strains and Uniform-Fields in Multiphase Elastic Media. *P Roy Soc Lond a Mat* 1992; 437: 291-310. DOI: DOI 10.1098/rspa.1992.0062.
6. Ammar K, Appolaire B, Cailletaud G, et al. Combining phase field approach and homogenization methods for modelling phase transformation in elastoplastic media. *European Journal of Computational Mechanics* 2009; 18: 485-523. DOI: 10.3166/ejcm.18.485-523.
7. Tirry W and Schryvers D. Linking a completely three-dimensional nanostrain to a structural transformation eigenstrain. *Nat Mater* 2009; 8: 752-757. DOI: 10.1038/Nmat2488.
8. Eshelby JD. Elastic inclusions and inhomogeneities. *Progress in solid mechanics* 1961; 2: 89-140.
9. Ma LF and Korsunsky AM. The principle of equivalent eigenstrain for inhomogeneous inclusion problems. *Int J Solids Struct* 2014; 51: 4477-4484.
10. Parnell WJ. The Eshelby, Hill, Moment and Concentration Tensors for Ellipsoidal Inhomogeneities in the Newtonian Potential Problem and Linear Elastostatics. *J Elasticity* 2016; 125: 231-294.
11. Zhong Z and Meguid SA. On the elastic field of a spherical inhomogeneity with an imperfectly bonded interface. *J Elasticity* 1997; 46: 91-113.
12. Lee S and Ryu S. Theoretical study of the effective modulus of a composite considering the orientation distribution of the fillers and the interfacial damage. *European Journal of Mechanics - A/Solids* 2018; 72: 79-87. DOI: https://doi.org/10.1016/j.euromechsol.2018.02.008.
13. Yanase K and Ju JW. Effective Elastic Moduli of Spherical Particle Reinforced Composites Containing Imperfect Interfaces. *Int J Damage Mech* 2012; 21: 97-127.
14. Qu J and Cherkaoui M. *Fundamentals of Micromechanics of Solidss*. John Wiley & Sons, Inc., 2007.
15. Qu JM. The Effect of Slightly Weakened Interfaces on the Overall Elastic Properties of Composite-Materials. *Mech Mater* 1993; 14: 269-281.
16. Jasiuk I and Tong Y. Effect of interface on the elastic stiffness of composites. In: *American Society of Mechanical Engineers, Applied Mechanics Division, AMD* 1989, pp.49-54.
17. Sohn D. Periodic mesh generation and homogenization of inclusion-reinforced composites using an element-carving technique with local mesh refinement. *Composite Structures* 2018; 185:

65-80. DOI: https://doi.org/10.1016/j.compstruct.2017.10.088.

18. Barai P and Weng GJ. A theory of plasticity for carbon nanotube reinforced composites. *International Journal of Plasticity* 2011; 27: 539-559. DOI: https://doi.org/10.1016/j.ijplas.2010.08.006.

19. Chiang CR. On Eshelby's tensor in transversely isotropic materials. *Acta Mech* 2017; 228: 1819-1833.

20. Jin XQ, Keer LM and Wang Q. A Closed-Form Solution for the Eshelby Tensor and the Elastic Field Outside an Elliptic Cylindrical Inclusion. *J Appl Mech-T Asme* 2011; 78.

21. Jin XQ, Lyu D, Zhang XN, et al. Explicit Analytical Solutions for a Complete Set of the Eshelby Tensors of an Ellipsoidal Inclusion. *J Appl Mech-T Asme* 2016; 83.

22. Wu TH, Chen TY and Weng CN. Green's functions and Eshelby tensors for an ellipsoidal inclusion in a non-centrosymmetric and anisotropic micropolar medium. *Int J Solids Struct* 2015; 64-65: 1-8.

23. Ma HM and Gao XL. Eshelby's tensors for plane strain and cylindrical inclusions based on a simplified strain gradient elasticity theory. *Acta Mech* 2010; 211: 115-129. DOI: 10.1007/s00707-009-0221-0.

24. Li S, Sauer RA and Wang G. The Eshelby Tensors in a Finite Spherical Domain—Part I: Theoretical Formulations. *Journal of Applied Mechanics* 2006; 74: 770-783. DOI: 10.1115/1.2711227.

25. Li S, Wang G and Sauer RA. The Eshelby Tensors in a Finite Spherical Domain—Part II: Applications to Homogenization. *Journal of Applied Mechanics* 2006; 74: 784-797. DOI: 10.1115/1.2711228.

26. Zou WN and He QC. Eshelby's problem of a spherical inclusion eccentrically embedded in a finite spherical body. *Proceedings of the Royal Society A: Mathematical, Physical and Engineering Science* 2017; 473. 10.1098/rspa.2016.0808.

27. Gao XL and Ma HM. Solution of Eshelby's inclusion problem with a bounded domain and Eshelby's tensor for a spherical inclusion in a finite spherical matrix based on a simplified strain gradient elasticity theory. *J Mech Phys Solids* 2010; 58: 779-797. DOI: https://doi.org/10.1016/j.jmps.2010.01.006.

28. Mejak G. Eshebly tensors for a finite spherical domain with an axisymmetric inclusion. *European Journal of Mechanics - A/Solids* 2011; 30: 477-490. DOI: https://doi.org/10.1016/j.euromechsol.2011.02.001.

29. Dekkers MEJ and Heikens D. The Effect of Interfacial Adhesion on the Tensile Behavior of Polystyrene Glass-Bead Composites. *J Appl Polym Sci* 1983; 28: 3809-3815.

30. Zhao FM and Takeda N. Effect of interfacial adhesion and statistical fiber strength on tensile strength of unidirectional glass fiber/epoxy composites. Part. II: comparison with prediction. *Compos Part a-Appl S* 2000; 31: 1215-1224.

31. Qu JM. Eshelby Tensor for an Elastic Inclusion with Slightly Weakened Interface. *J Appl Mech-T Asme* 1993; 60: 1048-1050.

32. Achenbach JD and Zhu H. Effect of Interfacial Zone on Mechanical-Behavior and Failure of Fiber-Reinforced Composites. *J Mech Phys Solids* 1989; 37: 381-393.

33. Wang Z, Zhu J, Chen WQ, et al. Modified Eshelby tensor for an ellipsoidal inclusion imperfectly embedded in an infinite piezoelectric medium. *Mech Mater* 2014; 74: 56-66.

34. Othmani Y, Delannay L and Doghri I. Equivalent inclusion solution adapted to particle debonding with a non-linear cohesive law. *Int J Solids Struct* 2011; 48: 3326-3335.

35. Folland GB. *Real analysis: modern techniques and their applications*. John Wiley & Sons, 2013.

36. Taylor AE. General theory of functions and integration. 1965.

37. Gupta S, Schieber JD and Venerus DC. Anisotropic thermal conduction in polymer melts in uniaxial elongation flows. *J Rheol* 2013; 57: 427-439.

38. Hagihara K, Kinoshita A, Sugino Y, et al. Plastic deformation behavior of Mg89Zn4Y7 extruded alloy composed of long-period stacking ordered phase. *Intermetallics* 2010; 18: 1079-1085.

39. Delaire F, Raphanel JL and Rey C. Plastic heterogeneities of a copper multicrystal deformed in uniaxial tension: experimental study and finite element simulations. *Acta Materialia* 2000; 48: 1075-1087. DOI: https://doi.org/10.1016/S1359-6454(99)00408-5.

40. Gavazzi AC and Lagoudas DC. On the numerical evaluation of Eshelby's tensor and its application to elastoplastic fibrous composites. *Comput Mech* 1990; 7: 13-19. DOI: 10.1007/BF00370053.

41. Giraud A, Huynh QV, Hoxha D, et al. Application of results on Eshelby tensor to the determination of effective poroelastic properties of anisotropic rocks-like composites. *Int J Solids Struct* 2007; 44: 3756-3772.

42. Dederichs PH and Leibfried G. Elastic Green's Function for Anisotropic Cubic Crystals. *Physical Review* 1969; 188: 1175-1183. DOI: 10.1103/PhysRev.188.1175.

43. Yu HY, Sanday SC and Chang CI. Elastic Inclusions and Inhomogeneities in Transversely Isotropic Solids. *P R Soc-Math Phys Sc* 1994; 444: 239-252.

44. Dunn ML and Wienecke HA. Inclusions and inhomogeneities in transversely isotropic piezoelectric solids. *Int J Solids Struct* 1997; 34: 3571-3582.

45. Hashin Z. The Spherical Inclusion with Imperfect Interface. *J Appl Mech-T Asme* 1991; 58: 444-449.

46. Walpole LJ. A coated inclusion in an elastic medium. *Mathematical Proceedings of the Cambridge Philosophical Society* 1978; 83: 495-506. 2008/10/24. DOI: 10.1017/S0305004100054773.

47. Qiu YP and Weng GJ. Elastic Moduli of Thickly Coated Particle and Fiber-Reinforced Composites. *Journal of Applied Mechanics* 1991; 58: 388-398. DOI: 10.1115/1.2897198.

48. Mikata Y and Taya M. Stress Field in and Around a Coated Short Fiber in an Infinite Matrix Subjected to Uniaxial and Biaxial Loadings. *Journal of Applied Mechanics* 1985; 52: 19-24. DOI: 10.1115/1.3168996.

49. Duan HL, Wang J, Huang ZP, et al. Stress concentration tensors of inhomogeneities with interface effects. *Mech Mater* 2005; 37: 723-736.

50. Duan HL, Wang J, Huang ZP, et al. Size-dependent effective elastic constants of solids containing nano-inhomogeneities with interface stress. *J Mech Phys Solids* 2005; 53: 1574-1596.

51. Duan HL, Wang J, Huang ZP, et al. Eshelby formalism for nano-inhomogeneities. *P Roy Soc a-Math Phy* 2005; 461: 3335-3353.

52. Luding S. Cohesive, frictional powders: contact models for tension. *Granular Matter* 2008; 10: 235. DOI: 10.1007/s10035-008-0099-x.

53. Comsol AB. COMSOL MultiPhysics Reference Manual, version 5.3. 2015.

54. Brown JM, Abramson EH and Angel RJ. Triclinic elastic constants for low albite. *Phys Chem Miner* 2006; 33: 256-265.

55. Qiu YP and Weng GJ. On the application of Mori-Tanaka's theory involving transversely isotropic spheroidal inclusions. *International Journal of Engineering Science* 1990; 28: 1121-1137. DOI: https://doi.org/10.1016/0020-7225(90)90112-V.

56. Helnwein P. Some remarks on the compressed matrix representation of symmetric second-order and fourth-order tensors. *Comput Method Appl M* 2001; 190: 2753-2770.

57. Voigt W. *Lehrbuch der kristallphysik (mit ausschluss der kristalloptik)*. Springer-Verlag, 2014.

58. Grytten F, Holmedal B, Hopperstad OS, et al. Evaluation of identification methods for YLD2004-18p. *International Journal of Plasticity* 2008; 24: 2248-2277. DOI: https://doi.org/10.1016/j.ijplas.2007.11.005.

59. Messner M, Beaudoin A and Dodds RH. A grain boundary damage model for delamination. *Comput Mech* 2015; 56: 153-172.

60. Yan W, Lin S, Kafka OL, et al. Data-driven multi-scale multi-physics models to derive process–structure–property relationships for additive manufacturing. *Comput Mech* 2018; 61: 521-541. DOI: 10.1007/s00466-018-1539-z.

61. Milgrom M and Shtrikman S. The energy of inclusions in linear media exact shape-independent relations. *J Mech Phys Solids* 1992; 40: 927-937. DOI: https://doi.org/10.1016/0022-5096(92)90056-8.

62. Kuppers H and Siegert H. Elastic Constants of Triclinic Crystals Ammonium and Potassium Tetroxalate Dihydrate. *Acta Crystall a-Crys* 1970; A 26: 401-&.

63. Sedlak P, Seiner H, Zidek J, et al. Determination of All 21 Independent Elastic Coefficients of Generally Anisotropic Solids by Resonant Ultrasound Spectroscopy: Benchmark Examples. *Exp Mech* 2014; 54: 1073-1085.